\documentclass[nofootinbib,10pt,preprintnumbers]{revtex4}
\usepackage[T1]{fontenc}
\usepackage[latin1]{inputenc}
\usepackage{graphicx}
\usepackage{amsfonts}
\usepackage{amssymb}
\usepackage{calc}
\usepackage{ifthen}
\usepackage{color}

{\newcommand{\lsim}{\mbox{\raisebox{-.6ex}{~$\stackrel{<}{\sim}$~}}}
{\newcommand{\gsim}{\mbox{\raisebox{-.6ex}{~$\stackrel{>}{\sim}$~}}} 
{
\newcommand{\be}{\begin{equation}}
\newcommand{\ee}{\end{equation}}
\newcommand{\bea}{\begin{eqnarray}}
\newcommand{\eea}{\end{eqnarray}}

\newcommand{\lam}{\lambda}
\renewcommand{\epsilon}{\varepsilon}

\newcommand{\bean}{\begin{eqnarray*}}
\newcommand{\eean}{\end{eqnarray*}}

\def\GeV{{\rm \ GeV}}
\def\MeV{{\rm \ MeV}}

\def\TeV{{\rm \ TeV}}

\begin{document}
\preprint{ULB-TH/10-04}
\title{A Tight Connection Between Direct and Indirect Detection of Dark Matter\\ 
through Higgs Portal Couplings to a Hidden Sector }
\author{Chiara Arina}
\email{carina@ulb.ac.be}
\author{Fran\c{c}ois-Xavier Josse-Michaux}
\email{fxjossemichaux@gmail.com}
\author{Narendra Sahu}
\email{Narendra.Sahu@ulb.ac.be}
\affiliation{Service de Physique Th\'eorique, Universit\'e Libre de Bruxelles, 1050 Brussels,
Belgium}
\begin{abstract}
We present a hidden abelian extension of the Standard Model including a complex scalar as a dark matter candidate and a light scalar acting as a long range force carrier between dark matter particles. The Sommerfeld enhanced annihilation cross-section of the dark matter explains the observed cosmic ray excesses. The light scalar field also gives rise to potentially large cross-sections of dark matter on the nucleon, therefore providing an interesting
way to probe this model simultaneously at direct and indirect dark matter search experiments. We constrain the parameter space of the model by taking into account CDMS-II exclusion limit as well as PAMELA and FermiLAT data. 
\end{abstract}

\pacs{95.35.+d}

\maketitle

\section{Introduction}
The existence of dark matter (DM) in the present Universe has been firmly supported 
by a range of evidences~\cite{dm_review}. The prime among them are galaxy rotation curves, 
large scale structure, cosmic microwave background and gravitational lensing. However, the 
identity of DM within the Standard Model (SM) of particle physics is still missing. 
Therefore, its experimental verification is expected to be a new discovery and a strong indication 
for physics beyond the SM. 

Recently the Cryogenic DM Search (CDMS) Collaboration~\cite{cdms} in the Soudan mine reported the observation 
of two events compatible with a positive dark matter detection at $1.64\  \sigma$ confidence level (C.L.). The DAMA/LIBRA~\cite{dama} experiment claims an 
evidence of DM in its modulated signal at $8.8\ \sigma$ C.L. Several other direct detection experiments 
are running and setting upper bounds, including EDELWEISS-II~\cite{edelweiss}, ZEPLIN-II~\cite{zeplin}, 
XENON-10~\cite{xenon2007} and Cresst-III~\cite{cresst}. The Xenon collaboration will soon release the data 
from the first run of the Xenon100 experiment~\cite{xenonup} and superCDMS is planned. 

A significant amount of effort has also been devoted to detection of DM through indirect searches. For example, the 
PAMELA collaboration~\cite{pamela} reported an unexpected rise of the positron fraction compared to that of 
the galactic background at energies above 10 GeV, while confirming the earlier results of AMS~\cite{ams} 
and HEAT~\cite{heat}. Similarly the HESS~\cite{hess} and Fermi Large Area Telescope (FermiLAT)~\cite{fermilat}
collaborations also reported an excess of electron plus positron flux with respect 
to the galactic background at energies above 100 GeV, but without confirming the spectral features 
observed by the balloon-based experiments ATIC~\cite{atic} and PPB-BETS~\cite{ppbbets}. It has been 
widely interpreted that DM could be a viable candidate for the observed cosmic ray anomalies, although they could be 
explained by astrophysical sources~\cite{astrophysics}.   

In light of the above experimental results, several models have been considered in the literature, which leave signatures at either direct\cite{direct_detection} or indirect \cite{indirect_scattering,indirect_decay} DM searches. Typically, for a given model, the predictions for direct and indirect signatures of DM depend on different parts of the parameter-space, and the derived constraints thus do not overlap.
However, in some models the same couplings are responsible for both the scattering of DM on the nucleon and large annihilation cross sections, in which case an interesting complementarity between direct and indirect searches exists~\cite{cfs,cfs2}.

A simple possibility is to consider  
singlet extensions of the 
SM, in which the DM is a singlet scalar~\cite{Singlet}, coupled to 
the SM Higgs particle through the so-called Higgs portal~\cite{wilzeck}, namely the Higgs to DM quartic coupling (see also doublet extensions of the SM~\cite{doublet_models}). Depending 
on the strength of this portal, the singlet can account for the observed relic density, 
$\Omega_{\rm DM} h^{2}\sim 0.1$~\cite{Komatsu:2010fb}. Furthermore, through the Higgs portal coupling, DM scatters with the 
nucleon and is thereby constrained by direct searches, as well as annihilates into SM fermions which can be observed at indirect detection experiments~\cite{cfs,cfs2}. 

The anomalous 
positron and electron fluxes observed by PAMELA and FermiLAT
require a large enhancement of current DM annihilations. 
In their minimal versions~\cite{Singlet}, the singlet extensions cannot reproduce such features. Furthermore, in these models, the stability of DM is supported by an {\textit{ad hoc}} discrete symmetry.
In this paper, we study a model which naturally solves these issues. 
We introduce a hidden sector gauged under an Abelian $U(1)_{H}~$\cite{HiddenU1}-\cite{HU1_2}, containing two complex scalars $\Phi$ and $S$. 
While all SM fields are hidden sector singlet, the extra scalars are singlet under the SM but charged under $U(1)_{H}$. This model provides 
all the ingredients for a viable DM model with potentially large direct and indirect detection signals.

The paper is organized as follows. In Sec.~\ref{sec:model} we present the model, with particular attention to the mass spectrum and mixings in the scalar sector. The model parameter space is then examined in 
Sec.~\ref{sec:omegah2} and constrained by requiring the relic density of the DM candidate to be in the WMAP7 range. Sections~\ref{sec:dd} and \ref{sec:ind} describe 
the phenomenology of the model in light of the present DM searches: CDMS-II, XENON10, PAMELA and FemiLAT. First, in Sec.~\ref{sec:dd}, 
we discuss the direct detection bounds constraining the hidden sector parameter space and the interplay 
with the indirect detection. Then, in Sec.~\ref{sec:ind}, we investigate the indirect detection 
bounds, and give the results in terms of positron and electron excesses. The 
conclusions are presented in Sec.~\ref{sec:conclusion}.
\section{The Model} \label{sec:model}
\begin{figure}[h!]
\begin{center}
\includegraphics[scale=0.5]{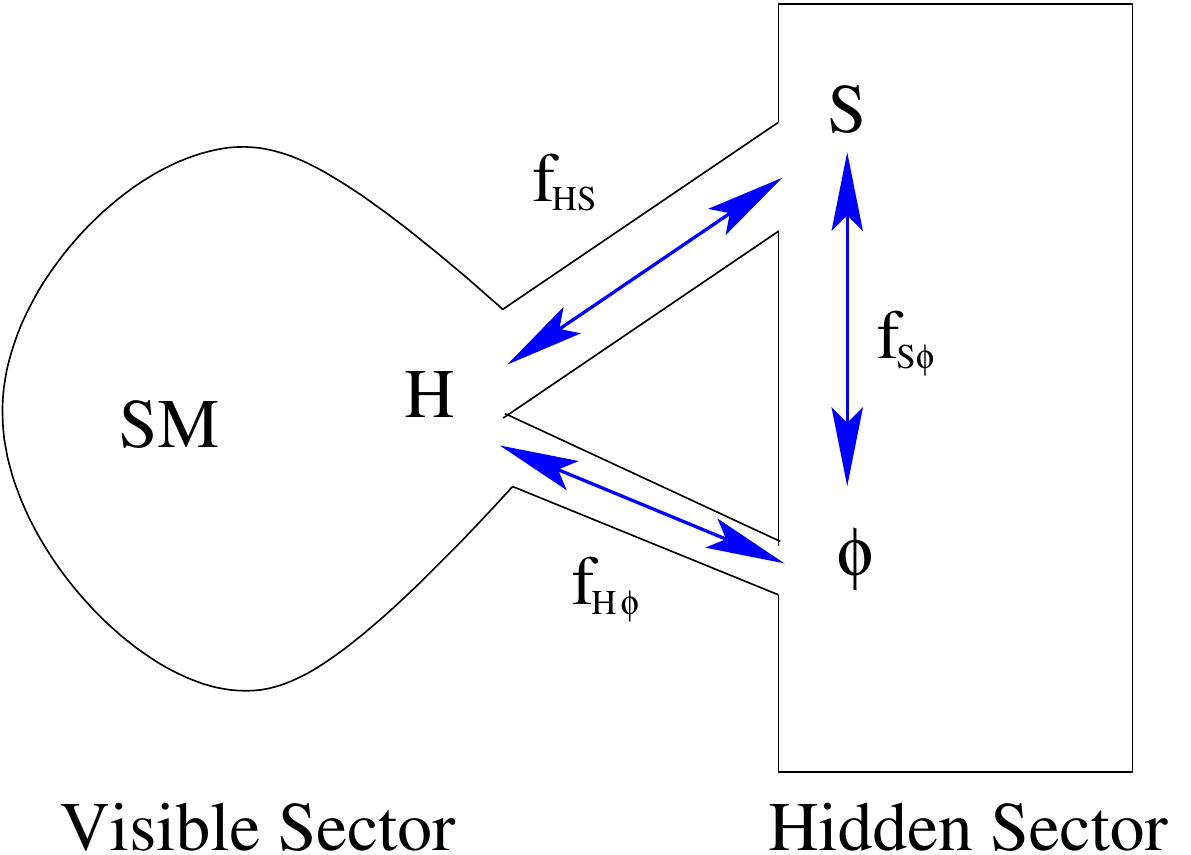}
\caption{Schematic representation of Higgs portal couplings to a hidden sector.}
\label{fig-1}
\end{center}
\end{figure}

Assuming a non-trivial charge assignment under the hidden Abelian $U(1)_{H}$ for the extra 
scalars $S$ and $\Phi$, $q_{H}(S)=3$ and $q_{H}(\Phi)=2$, the most general scalar potential 
is given by:
\bea
V(H,S,\Phi) & = & -\mu_H^2 H^\dagger H + \lambda_H (H^\dagger H)^2 + \mu_S^2 S^\dagger S 
+ \lambda_S (S^\dagger S)^2 - \mu_\phi^2 \Phi^\dagger \Phi + \lambda_\phi (\Phi^\dagger \Phi)^2 \nonumber\\
& & +f_{HS} H^\dagger H S^\dagger S +  f_{H\phi} H^\dagger H \Phi^\dagger \Phi 
+f_{S\phi} S^\dagger S \Phi^\dagger \Phi \,.
\label{scalar_potential}
\eea 
The hidden sector couples to the SM via Higgs portals, as schematically depicted in Fig.(\ref{fig-1}).
We assume that all parameters are real and positive.
The field $\Phi$ acquires a nonzero vacuum expectation value (vev), which triggers the breaking of $U(1)_{H}$ to a remnant $Z_{2}$ symmetry 
under which all fields are even but $S$.
The latter does not develop any nonzero vev and hence can be a dark matter candidate, being stabilized by the $Z_2$ symmetry.

It is remarkable that almost all the parameters in this model are well constrained by both direct and indirect searches, as we will see in great details below. 
The relic density of $S$ is obtained through its annihilations into $H$ and $\Phi$ via $f_{HS}$ and $f_{S\phi}$, respectively. These two couplings also enter in the 
spin-independent cross section of DM on nucleon and hence are strongly 
constrained by direct DM searches. The field $\Phi$ gives rise to a long range attractive force 
between $S$ particles, thus boosting the current $S$ annihilations, 
while keeping the relic abundance unchanged. As a result $f_{S\phi}$ gets 
strongly constrained from indirect DM searches. After $\Phi$ acquires a vev, it mixes 
with the SM Higgs through $f_{H\phi}$. $\Phi$ is destabilized and consequently decays 
into SM fermions through the same coupling. While $f_{H\phi}$ should not be too small for $\Phi$ to 
decay before the onset of big bang nucleosynthesis (BBN), a too large $H$-$\Phi$ mixing is excluded 
by constraints coming from LEP-II on the ratio of the invisible-to-visible Higgs decay~\cite{CBhiggs1}. 
\subsection{Masses and Mixings of Hidden Sector Fields}\label{sec:masses}
From the scalar potential in Eq.~(\ref{scalar_potential}), let us derive the quantities relevant for this 
study. First of all, vacuum stability requires, 
besides positive quartic couplings, that $4\lam_{H}\lam_{S}\lam_{\phi}+f_{HS}f_{H\phi}f_{S\phi} 
\geq \lam_{H}f_{S\phi}+\lam_{\phi}f_{H S}+\lam_{S}f_{H \phi}$. The electroweak symmetry breaking 
occurs when the SM Higgs acquires a vev $\langle H \rangle = v/\sqrt{2}$, while $U(1)_{H}$ is broken 
to a surviving $Z_2$ symmetry when $\Phi$ acquires a vev $\langle \Phi \rangle = u/\sqrt{2}$. In the 
unitary gauge, the quantum fluctuations around the minimum are parametrized as
\begin{equation}
H=\frac{1}{\sqrt{2}}\pmatrix{0\cr v+h}~~~{\rm and}~~~\Phi=\frac{1}{\sqrt{2}}(u + \phi)\,,
\label{q_fluct}
\end{equation}
where $h$ and $\phi$ are physical real scalars, the unphysical degrees of freedom being eaten by the 
longitudinal component of the SM gauge bosons and of the $Z'$ associated with $U(1)_{H}$.
Minimization of the scalar potential in Eq.~(\ref{scalar_potential}) enforces 
\begin{eqnarray}
\frac{v}{\sqrt{2}} = \left( \frac{ 2\mu_H^2\,\lambda_\phi -\mu_\phi^2\,f_{H\phi}  }
{4 \lambda_H \lambda_\phi -f_{H\phi}^2  } \right)^{1/2}\,,\quad 
\frac{u}{\sqrt{2}} = \left(\frac{2\mu_\phi^2\,\lambda_H- \mu_H^2\,f_{H\phi} }
{4\lambda_H \lambda_\phi - f_{H\phi}^2} \right)^{1/2}\,.
\end{eqnarray}
This minimum is the global one if $f_{H\phi}^{2}\leq 4 \lam_{H}\lam_{\phi}$. The two real scalars 
$h$ and $\phi$ mix with each other and the mass matrix in the basis spanned by $(h,\phi)$ is given by:
\begin{equation}
{\mathcal M}^2=\pmatrix{ 2 \lambda_H v^2 &   f_{H\phi} u v\cr
  f_{H\phi} u v &  2 \lambda_\phi u^2}\,.
\end{equation}
Assuming $f_{H\phi}\ll 1$, the mixing angle between $h$ and $\phi$ is suppressed:
\be
\theta_{m} \sim -\frac{f_{H\phi}\,u\,v}{2\left(\lambda_H v^2- \lambda_\phi u^2\right)}\ll 1\,.
\label{eq:theta}
\ee
In terms of $(h,\phi)$, the mass eigenstates $h_{1}$ and $h_{2}$ read:
\bea
h_{1} \sim  h + \theta_{m}\, \phi\,~~~{\rm and}~~~ h_{2} \sim  \phi -\theta_{m}\, h\,.
\eea
Consequently $h_{1}$ is mostly the SM Higgs field, while $h_{2}$ is the light scalar. Their respective masses are:
\begin{equation}
M_{h_{1}}^2 \simeq 2 \lambda_H v^2~~~{\rm and}~~~M_{h_2}^2 \simeq  2 \lambda_\phi u^{2}\left(1 - 
\frac{f_{H\phi}^2 }{4 \lambda_H \lambda_\phi} \right)\,.
\end{equation}
Because of the small mixing, the current experimental bounds~\cite{CBhiggs1,CBhiggs2} on the SM Higgs mass apply on $M_{h_{1}}$.
Hereafter, in all numerical evaluations we take $M_{h_{1}}=120\GeV$. However, we note that our results are 
quite insensitive to the Higgs Mass. The light scalar field $h_2$ is the main product of $S$ 
annihilations at the present epoch. In order to avoid overproduction of high-energy gamma rays from the 
decay of $h_2$, $M_{h_2}< 2 m_{\pi}$ is required: we take $M_{h_{2}}=240\MeV$ as numerical reference value. 
We then have $\lambda_H v^2 \gg \lambda_\phi u^{2}$.\\
The DM mass is given by:
\be
M_S^2=\mu_S^2 + \frac{f_{S\phi}}{2} u^{2} + \frac{f_{HS}}{2} v^{2} \,,
\ee
which is varied from $10 \GeV$ to $1\TeV$ in the following. 

The hidden sector is gauged under the Abelian $U(1)_{H}$ and contains an extra gauge 
boson $Z'$. Although SM particles are singlet under $U(1)_{H}$, through the kinetic mixing of $Z'$ with the hypercharge gauge boson, they can couple to $Z'$. The relevant part of the Lagrangian then reads:
\begin{equation}
\mathcal{L}_{Z'} = -\frac{1}{4}F^{H}_{\mu\nu}F^{H\,\mu\nu} -\frac{\chi}{2}F^Y_{\mu\nu}F^{H\,\mu\nu}
+ \big\vert(\partial_{\mu} - 2 i g_H Z'_{\mu})\Phi\big\vert^2 + \big\vert(\partial_{\mu} 
- 3 i g_H Z'_{\mu}) S\big\vert^2\,,
\label{l_hidden}
\end{equation}
where $g_H$ is the hidden sector coupling constant,  $F^Y_{\mu\nu}$ and $F^{H}_{\mu\nu}$ are the 
field strength tensors associated with $U(1)_{Y}$ and $U(1)_{H}$ respectively and $\chi$ parameterizes 
the kinetic mixing between the $U(1)$ symmetries. 

For an invisible $Z'$, the kinetic mixing is expected to be $\chi \sim 10^{-4}$-$10^{-2}$ at the electroweak 
scale. The mass of $Z'$ is given by $M_{Z'}= 2 g_H u$ and we take
600 GeV as a reference numerical value. The parameter $u$ is then lower bounded by 
the requirement of perturbative $U(1)_{H}$ gauge coupling: $u\gtrsim 85\GeV$. For the chosen value of the $Z'$ mass, the electroweak precision measurement constraints the kinetic 
mixing to be $\chi\leq 0.036$~\cite{kinmix}. Through $Z$-$Z'$ mixing, the DM can scatter on nuclei. The corresponding cross section may exceed the current experimental bounds for $\chi \gtrsim 
0.01$. When $M_{Z'}\lesssim M_{S}$ the dark gauge boson is produced in $S$ annihilations. For large $\chi$, this 
would lead to significant antiproton fluxes in the cosmic ray which is in contradiction with PAMELA data~\cite{pamela_antiproton}. In the following we assume $\chi \sim 10^{-4}$ and so $Z'$ effects are negligible.

Assuming $M_{Z'} = 600\GeV$ with $g_{H}\sim {\mathcal O}(1)$ while $M_{h_{2}}=240\MeV$ implies a strong tuning of the model parameters, with $\mu_{\phi}=120\MeV\ll u\sim 300 \GeV$ and $\lambda_{\phi}\sim 10^{-7}$. Reducing this tuning can be achieved by either increasing the hidden scalar boson mass, or by lowering $M_{Z'}$ as well as $M_{S}$. However, in those cases, predictions of the model turn out to be very different from those we discuss hereafter. By increasing $M_{h_{2}}$ we possibly face an antiproton overproduction in DM current annihilations, in which case the corresponding cross section should be suppressed compared to PAMELA requirements. In this case, still, the model can afford a viable DM candidate but lacks predictivity. Another way to reduce the tuning is to lower $M_{Z'}$ or $M_{S}$. With a lighter DM, the PAMELA cosmic ray spectrum cannot be accounted for, while with a lighter $Z'$ the model is in conflict with direct detection experiments. Therefore, in the following, we assume a certain amount of tuning, allowing the model to be predictive and probed by both direct and indirect detection experiments.

\subsection{Astrophysical and collider constraints}
We now discuss some relevant constraints on the hidden sector coming from astrophysics and the electroweak 
precision measurements. The scalar field $h_{2}$ decays 
to SM fermions through $\Phi$-$H$ mixing; its decay rate is approximately given by:
\begin{equation}
\Gamma_{h_2}\simeq \sum_{2m_{f} < M_{h_{2}}}\frac{M_{h_{2}}}{8\pi}\theta_{m}^{2}\frac{2m_{f}^{2}}{v^{2}}\,.
\end{equation}
The lifetime of $h_{2}$ is then estimated as
\bea
\tau_{h_{2}}\simeq 0.1 s \times \frac{2m_{\pi}}{M_{h_{2}}}\times\left(\frac{\theta_{m}}{10^{-7}}\right)^{-2}\,.
\eea
Thus, demanding that $h_{2}$ decays before the onset of BBN $\tau_{h_2} \lsim \tau_{BBN}\sim 0.1 {\rm s}$~\cite{BBN}, $\theta_{m}$ should be bigger than 
$10^{-7}$. Equivalently a lower bound on the Higgs portal coupling $f_{H\phi}$ is inferred:
\bea
f_{H\phi}\gtrsim 10^{-8}\times\left(\frac{M_{h_{1}}}{120\GeV}\right)^{2}\,\left(\frac{u}{600\GeV} 
\right)^{-1}\,.
\eea
Assuming that $h_2$ dominantly decays to $\mu^+\mu^-$, a strong constraint on $\theta_m$ can be 
obtained from the $B$-meson decay. From the branching ratio $BR\left(B\to \mu^{+}\mu^{-}X\right) < 3.2 \times 
10^{-4}$~\cite{HU1_2},\cite{pdg}, one infers an upper bound on $\theta_{m}$ to be $\vert \theta_{m}\vert \lesssim 
10^{-2}$~\cite{O'Connell:2006wi}. This in turn gives
\bea
f_{H\phi}\lesssim 10^{-3} \times\left(\frac{600\GeV}{u}\right)\left(\frac{M_{h_{1}}}{120\GeV}\right)^{2}\,.
\eea
Since $h_2$ is very light, the mixing angle never saturates the above bound. For the same reason, the electroweak 
precision measurements on the S, T and U parameters do not receive any significant contributions from $h_2$~\cite{ewpo}.
\section{Annihilation cross-section of the $S$ DM at freeze-out and at present epoch}\label{sec:omegah2}
\subsection{Relic density} 
\begin{figure}[t!]
\includegraphics[width=1\textwidth]{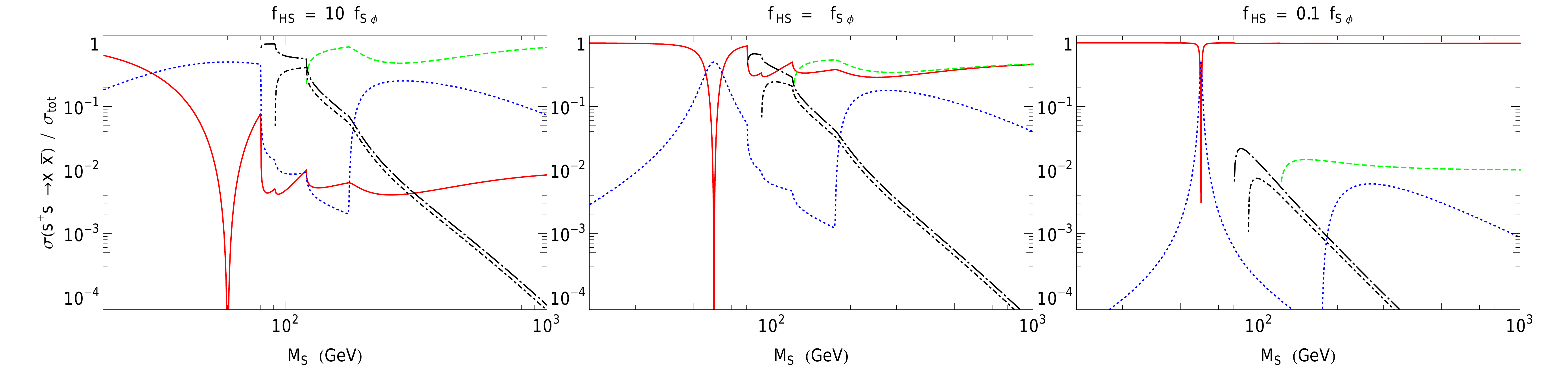}
\caption{Ratios of the leading annihilation channels of $S$ on the total annihilation cross-section are shown as a function of $M_S$:
in red solid $(S^{\dagger}S \to h_2 h_2)$, in green dashed $(S^{\dagger}S \rightarrow h_1 h_1)$, in blue dotted
$(S^{\dagger}S \rightarrow f\bar{f})$, in black dot-dashed $(S^{\dagger}S \rightarrow W^+ W^-)$ and in 
black dot-long dashed $(S^{\dagger}S \to Z Z)$.}
\label{fig-annchan}
\end{figure}

In the early universe the $S$ particles maintain their thermal equilibrium through the scattering 
processes: $S^\dagger S \to h_{2} h_{2}$ and $S^\dagger S \to$ SM particles. In Fig.~(\ref{fig-annchan}),  we display ratios of the dominant annihilation channels 
with respect to the total annihilation cross-section $\sigma_{\rm tot}$, as a function of the DM mass. In case of 
$f_{HS} >> f_{S\phi}$ (left panel) the dominant channel is $S^{\dagger}S \to \bar{f} f$ 
(blue dashed curve) until the $W^+ W^-$ (black dot-long dashed solid) and $Z Z$ (black dot-dashed) thresholds are reached. 
These two processes dominate the behavior of $S$ annihilations up to 120 GeV,  when $S^{\dagger}S\to h_1 h_1$ (green dashed curves) takes the upper hand. The situation is reversed as soon as $f_{S\phi} \gtrsim 
f_{HS}$ (central and right panels). The annihilation into $h_1 h_1$ is competitive only for equal Higgs 
portal couplings, while for $f_{S\phi} = 10 f_{HS}$, $S^{\dagger}S \rightarrow 
h_2 h_2$ (red solid curves) dominates in all the DM mass range, apart from the Higgs pole.

As the universe expands, the temperature of the thermal bath gradually falls; at $T_D \simeq M_S/25$, $S$ gets decoupled and starts redshifting. The relic 
abundance of DM can be evaluated by solving the Boltzmann equation for the $S$ number density:
\begin{equation}
\frac{d\,n_S}{dt} + 3 n_S H(t) = -\langle \sigma |v_{\rm rel}| \rangle \left( n_S^2 - 
{n_S^{\rm eq}}^2 \right)\,,
\label{boltzmann_equation}
\end{equation} 
where $H(t)$ is the Hubble expansion parameter and $\langle \sigma |v_{\rm rel}| \rangle$ is the thermal average of the total annihilation cross-section.   

To accurately evaluate the DM relic density, we solve Eq.(\ref{boltzmann_equation}) by using 
micrOMEGAs~\cite{micromegas}. The model has been implemented in micrOMEGAs through FeynRules~\cite{FeynRules}. The parameter space is chosen as follows. The physical Higgs masses are fixed at $M_{h_{1}}=120 \GeV$ and 
$M_{h_{2}}=240 \MeV$, while the other relevant parameters are varying randomly within their 
allowed ranges: $u$ from $10\GeV$ to $1\TeV$ and 
$\mu_{s}$ from $1\GeV$ to $500\GeV$. The portal couplings $f_{S\phi}$ and $f_{HS}$ vary 
from $10^{-3}$ to $\sqrt{4\pi}$, the perturbative upper bound, while $f_{H\phi}$ varies from 
$10^{-7}$ to $10^{-4}$. We also constrain the lifetime of $h_{2}$ to be less than $0.1$s. 

\begin{figure}[htb]
\begin{center}
\includegraphics[width=0.48\textwidth]{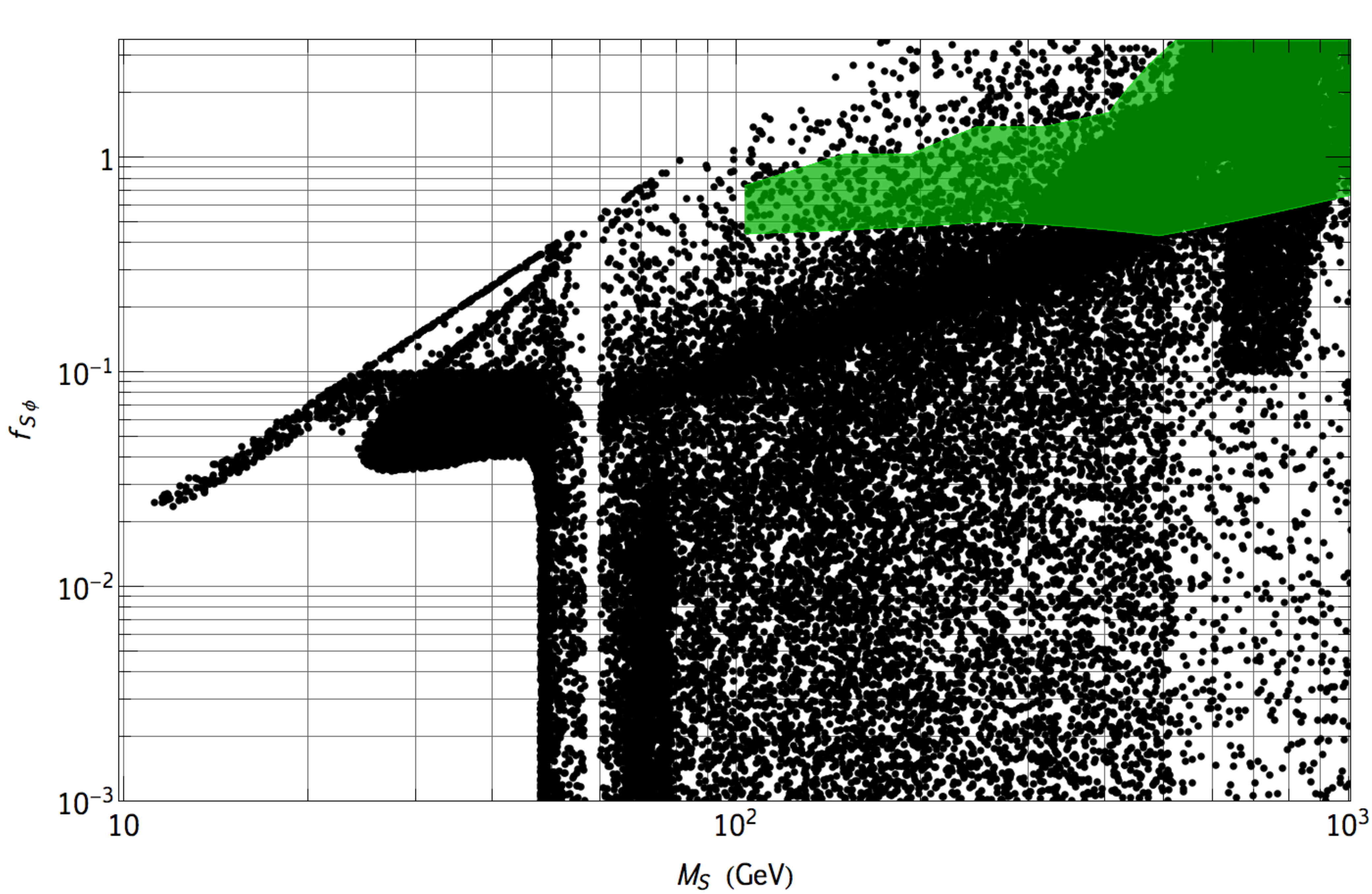}\quad \includegraphics[width=0.48\textwidth]
{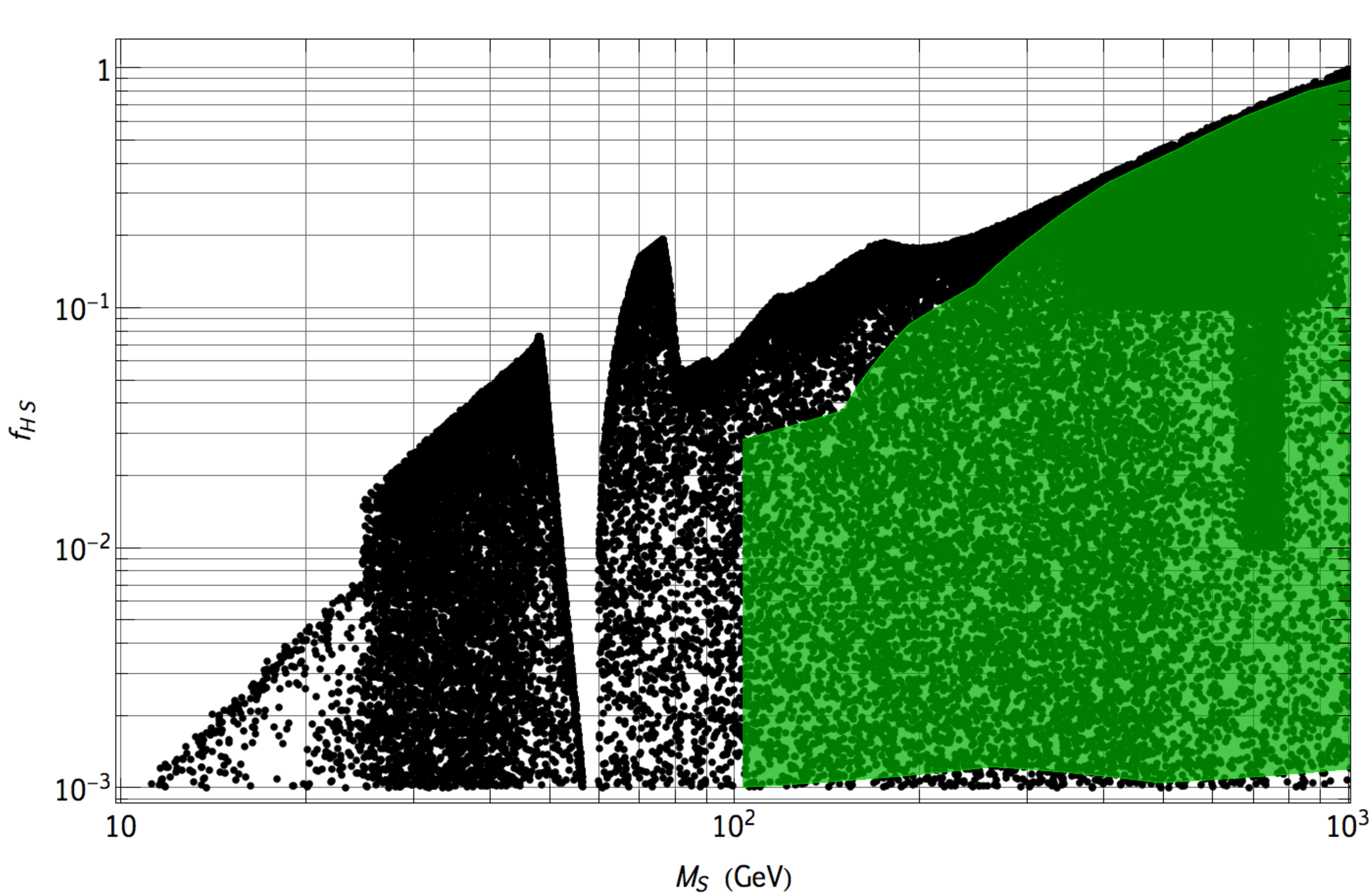}
\caption{Scatter plot for the relic abundance of $S$. In the left (right) panel the solutions that 
satisfy $\Omega_{\rm DM} h^2=0.0941-0.1277$ are shown as black points in the plane  $M_S$ versus $f_{S\phi}$ 
($f_{HS}$). The green (gray) region denotes points with boost factors compatible with indirect DM searches.}
\label{fig-2}
\end{center}
\end{figure}

Demanding that the relic abundance of DM 
should satisfy the WMAP7 constraints~\cite{Komatsu:2010fb} at $3\sigma$ C.L., given by 
\begin{equation}
\Omega_{\rm DM} h^{2}=0.0941-0.1277\,,
\end{equation} 
we show in Fig.~(\ref{fig-2}) the resulting scatter plots in the plane of $f_{S\phi}$ and $f_{HS}$ against $M_{S}$.

The requirement of having the correct relic density fixes the balance between the two 
Higgs portal couplings $f_{S\phi}$ and $f_{HS}$  and hence between the different annihilation 
channels. In the left panel of Fig.~(\ref{fig-2}), as long as $M_S\lesssim M_{h_{1}}/2$, we clearly see that the $S^\dagger S\to h_{2}h_{2}$ channel dominates. Near the Higgs resonance, $M_S \sim M_{h_{1}}/2$, lower values of 
the portal couplings are allowed, but at the pole, annihilations are too efficient and 
the relic density gets suppressed. For $M_{h_1}/2 \lesssim M_S \lesssim M_W$, the 
DM dominantly annihilates into light SM fermions, mostly $b \overline{b}$ pairs.
For $M_W \lesssim M_S \lesssim M_{h_{1}}$, the $S^{\dagger}S\to W^{+}W^{-}$ and $S^{\dagger}S\to ZZ$ channels, mediated by $h_{1}$, strongly constrain 
$f_{HS}$, as can be seen in Fig.~(\ref{fig-2}) (right panel).
For larger DM masses, 
the annihilation channels $S^{\dagger}S\to h_{1}h_{1}$ and $S^{\dagger}S\to \bar{t} t$ 
are also allowed. For such large DM masses, $f_{HS}$ is more constrained than $f_{S\phi}$. 
The latter can indeed take values up to the perturbative bound, while the former is upper bounded at $1$.

Resuming, on the whole range of DM mass, $f_{S\phi}$ typically takes values of the same order or larger than $f_{HS}$. This corresponds to middle and right plots of Fig.(\ref{fig-annchan}), and implies that the DM is 
preferably annihilating into $h_2$, as can be inferred from the approximate cross-sections:
\bea
\label{CrossSections}
\langle \sigma \vert v_{\rm rel}\vert \rangle(S^\dagger S \to h_2h_2) &\simeq&
\frac{f_{S\phi}^{2}}{64 \pi \,M_{S}^{2}}\,,\nonumber \\ \langle \sigma | v_{
\rm rel} |\rangle (S^\dagger S \to h_1h_1) &\simeq &\frac{1}{64 \pi}
\frac{f_{HS}^2}{M_S^2} \left( 1 -\frac{M_{h_{1}}^2}{M_S^2} \right)^{1/2}\,.
\eea   

Thus we see that both portal couplings $f_{HS}$ and $f_{S\phi}$ are constrained from the 
requirement of obtaining the right DM relic density. Interestingly, these couplings also enter in 
the t-channel Higgs-mediated scattering $S f\to S f$, which is relevant for the direct detection 
of DM, as we discuss in the next section. 

\subsection{Sommerfeld enhancement}
The observed relic density of DM can be obtained for an annihilation cross-section 
$\left\langle  \sigma|v_{\rm rel}| \right \rangle_{D} \simeq 3\times 10^{-26}{\rm cm}^{3}{\rm s}^{-1}$ 
at the decoupling epoch, $T_D \sim M_S/25$. However, at the present epoch, such a cross-section is 
too small to explain the anomalous cosmic ray fluxes observed by PAMELA and FermiLAT. 
Several mechanisms have been invoked in the literature to boost it. A natural 
enhancement may occur due to the local overdensities of DM in clumps~\cite{Lavalle:1900wn}, but 
estimates show that the resulting boost of the annihilation cross-section is too small to account 
for the observed cosmic ray fluxes. Two different mechanisms, arising from particle physics, also exist: either a Breit-Wigner resonance of 
the annihilation cross-section~\cite{HU1_2},\cite{BreitWigner}, or the Sommerfeld effect~\cite{Sommerfeld}. In this model, the latter naturally occurs because of the presence of $h_2$.
The light scalar $h_2$ acts as a long range attractive force carrier between the DM particles. If $S$ kinetic energy is small enough, the attractive interaction becomes 
relevant and induces an enhancement of the annihilation cross section. Defining the reduced DM bound-state wave-function as $\psi$, the corresponding boost is computed by solving the radial Schr\"{o}dinger equation:
\begin{equation}
\psi(r)^{\prime \prime}-M_{S}\,V(r)\psi(r)+M_{S}^2\,\beta^{2}\psi(r)\,=0,
\label{eqBF}
\end{equation}
where $\beta=v_{\rm rel}/c$ is the DM relative velocity, and $V(r)$ is the attractive Yukawa potential:
\begin{equation}
\label{SomPot}
V(r)= -\frac{\alpha}{r}\, e^{-M_{h_{2}}\,r}\,,\quad {\rm with}\quad \alpha = \left(\frac{f_{S \phi}  u}{M_{S}}\right)^{2} \frac{1}{8 \pi}\,.
\end{equation} 
We solve Eq.(\ref{eqBF}) using the boundary condition 
$\psi^{\prime}(\infty)={\rm  i} M_{S}\beta \,\psi(\infty)$.
The Sommerfeld boost is then given by 
\begin{equation}
S_{e}=\Big\vert \frac{\psi(\infty)}{\psi(0)}\Big\vert^{2}\,.
\end{equation}
To consider the enhancement at present time, one should integrate over the velocity distribution of DM in the Earth's neighborhood: 
\begin{equation}
\label{IntSom}
\langle S_e \rangle = \mathcal{N}_{\rm norm} \int_0^{v_{\rm esc}} d 
v_{\rm rel} \,\frac{v_{\rm rel}^2}{v_0^3} \, {\rm e}^{- v_{\rm rel}^2/v_0^2} 
S_e(v_{\rm rel},\alpha, M_{h_2}/M_S)\,,
\end{equation}
with mean velocity $v_0=220$ km/s, escape velocity $v_{\rm esc}= 650$ km/s and a normalization factor $\mathcal{N}_{\rm norm}$ for a smooth maxwellian halo.
The boost factor is then only a function of $u\times f_{S\phi}$, $M_{S}$ and $M_{h_2}$. Therefore, for a given DM mass, $\langle S_e \rangle$ is degenerate with respect to $f_{S\phi}$ and $u$, the hidden sector breaking scale.
The effective DM annihilation cross-section then reads 
\begin{equation}
\left\langle  \sigma |v_{\rm rel}| \right \rangle = \langle S_{e} \rangle \times \left\langle  \sigma|v_{\rm rel}| 
\right \rangle_{D}\,.
\end{equation} 

In Fig.~(\ref{fig-2}) the region of the parameters giving rise to boost factors from 20 up to $\sim$2000 is 
displayed by a green (gray) region. This is the range of enhancement required to explain 
the observed cosmic ray anomalies for a DM of mass (100 - 1000) GeV, compatible with current cosmological and astrophysical constraints, as discussed in Sec.~\ref{sec:ind}. On the left panel we see that 
only large values for $f_{S\phi}$ are allowed, ranging from $10^{-1}$ up to $2\sqrt{\pi}$. This is 
expected by the behavior of the Sommerfeld enhancement $\propto \alpha/\beta$.
As mentioned before, the requirement of having a good relic density fixes the balance between the two portal couplings. In the right panel of Fig.(\ref{fig-2}), the values of $f_{HS}$ in the green (gray) region correspond to the values of $f_{S\phi}$ giving rise to a correct boost factor, although the Sommerfeld effect does not depend on $f_{HS}$. For $100\GeV \lesssim M_{S}\lesssim 200\GeV$, the largest values of $f_{HS}$ are not allowed because 
they dominate $S$ annihilations, which in turn corresponds to lower values of $f_{S\phi}$. 
\section{CDMS-II Events and $S$ Dark Matter}\label{sec:dd}
\subsection{Experimental upper bounds}
The CDMS collaboration has recently published the analysis of the final run of the CDMS-II 
experiment~\cite{cdms}. After the background subtraction and cuts, two events survive, respectively 
at 12.3 keV and 15.5 keV recoil energies. The significance of the two events being a DM signal is at 
$1.64\sigma$ C.L., namely there is $23\%$ probability that these are of more common 
origin, such as cosmogenic or neutron background. In Ref.~\cite{cdmsanalysis}, it has been shown that if the 
two events are taken to be DM signal, the $1.64\sigma$ region will prefer light dark matter candidates, with an 
upper bound on the mass around 60-80 GeV. If we consider the 90$\%$ C.L. region to constraint 
the DM mass, then only an upper bound can be set. We approach the analysis of the two events of the 
CDMS-II in a conservative way, finding a 90$\%$ C.L. exclusion limit with the maximum gap method~\cite{mgm}. 

In addition the upper bound from the XENON10 experiment~\cite{xenon2007} is taken into account, which is the most 
constraining one together with CDMS-II, in the case of spin-independent elastic interaction on nucleon. 
Using the data of the 2007 run, reanalyzed as in Ref.~\cite{xenon2009}, and the maximum gap method, we 
infer the $90\%$ C.L. upper bound. We also consider the prediction for the first run of the XENON100 experiment, 
following Ref.~\cite{xenonup}.
\subsection{The elastic spin-independent $S$ cross-section}

\begin{figure}[t!]
 \includegraphics[width=0.6\textwidth]{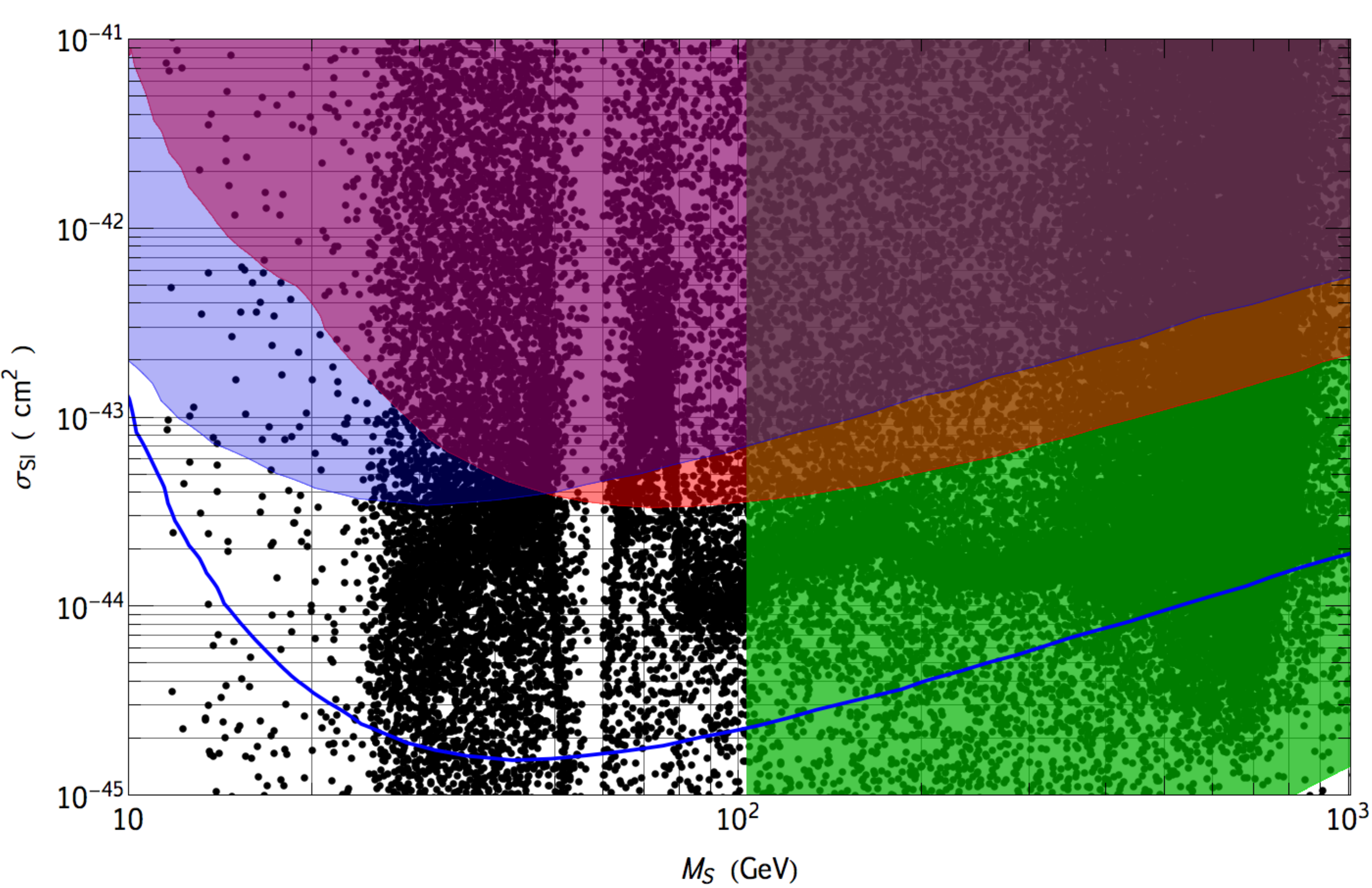} 
  \caption{Scatter plot for the direct detection of $S$ particles in the plane $M_S$ versus 
$\sigma_{\rm \rm SI}$. Solutions with a relic density in the WMAP7 range are depicted by black points; the 
red region (dark grey) is excluded at $90\%$ C.L. by CDMS-II, while the blue region (light gray) is 
excluded at $90\%$ C.L. by XENON10. The solid blue curve is the forecast upper bound for the Xenon100 
experiment. The green region (gray) denotes points that in addition provide a correct boost factor.}
\label{fig-5}
\end{figure}

The interaction of $S$ on the nucleon gives rise to a coherent spin-independent elastic 
scattering, mediated at tree level by the scalars of the model, the SM Higgs particle and $h_2$. 
This cross section reads:
\begin{equation}
\sigma_{\rm \rm SI} = \frac{\mu_{n}^2}{\pi\,  M^2_S} m^2_{n} f_{n}^2 \Big|\frac{1}{2}
\frac{f_{HS}}{M^2_{h_1}}+\frac{1}{2}\frac{f_{S\phi}\, u\, \theta_{m}}{M^2_{h_2} v} \Big|^2\,,
\label{eq:sigmadd}
\end{equation}
where $m_{n}$ is the nucleon mass and $\mu_{n}$ is the nucleon-DM reduced mass.
The parameter $f_{n}$ indicates the effective Higgs nucleon interaction, $f_{n} m_{n}= \langle {n} 
\vert \sum_q m_q \bar{q}q \vert {n}\rangle$, where the sum runs over all the quark flavor.
The $f_{n}$ factor introduces hadronic uncertainties in the elastic cross section: its value vary 
within a wide range $0.14 < f_{n} < 0.66$, as quoted in Refs.~\cite{fn_factor}. Hereafter we take 
$f_n$ to be 1/3, the central value.

From the Lagrangian in Eq.~(\ref{l_hidden}), the $S$ particle couples directly to the hidden gauge boson
 and through the kinetic mixing it communicates to the SM fermions. This results in an additional DM-nucleon scattering. Neglecting terms of order $\chi^3$, the 
corresponding spin-independent cross section is approximately:
\begin{equation}
\sigma^{\rm SI}_{n} (S {n} \rightarrow Z' \rightarrow S {n}) \simeq \frac{\mu_{n}^2}{A^2} 
\frac{g^2}{64 \pi \cos^2\theta_W}\frac{9}{M^2_{Z'} u^2} \chi^2 \sin^2\theta_W (A+2Z)^2\,,
\end{equation}
where $A$ and $Z$ are the mass and the atomic number of the nucleus and $g$ is the $SU(2)_{\rm L}$ 
coupling. The cross section is slightly dependent on the DM mass, through the reduced 
mass $\mu_{n}$. For  $M_{Z'}=600\GeV$, the values of $u$ we scanned over and for a maximal 
kinetic mixing $\chi=0.036$, the cross-section value ranges from $10^{-45}$cm$^{2}$ up to $10^{-43}$cm$^{2}$. 
As stated above, we fixed $\chi=10^{-4}$, and consequently the cross section varies from 
$10^{-50}$cm$^{2}$ up to $10^{-48}$cm$^{2}$, and is therefore a negligible correction to $h_{1}$ and $h_{2}$ contributions. 
It is then possible to constrain Higgs portal couplings thanks to direct detection searches.

The behavior of $\sigma_{\rm SI}$ as a function of the Higgs portal couplings is rather involved. 
Owing to the large mass differences between the two scalars, \textit{cf.} Eq.~(\ref{eq:theta}), the mixing angle between $h_2$ and $h_1$ is small, of the order of $\theta_{m} \sim \mathcal{O} (10^{-6}$-$10^{-5})$. Even though $\theta_{m}\ll1$, the second term in Eq.~(\ref{eq:sigmadd}) is not 
negligible in the whole $S$ mass range. With respect to the standard Higgs exchange, the $h_2$ 
contribution is enhanced due to its small mass, which compensates the smallness of its coupling 
to the nucleon. Note that this result is valid in general in models where a scalar with mass lighter than $1\GeV$, typically the light force carrier of the Sommerfeld enhancement mechanism, mixes with the SM Higgs~\cite{cfs,cfs2}.
  
The predictions for $\sigma_{\rm SI}$ are shown in Fig.~(\ref{fig-5}) as a function of $M_S$, with all 
the points having a relic abundance in the WMAP7 range. The cross-section is enhanced respect to the 
standard Higgs exchange: a large region (red region) of the parameter space is not allowed by 
CDMS-II experiment. In the low mass range a portion of the parameter space is incompatible with the 
XENON10 upper bound, denoted by the blue (light gray) band. The green (gray) region describes the parameter space that leads 
to a large sommerfeld effect and boosts $\langle\sigma |v_{\rm rel}|\rangle_{D}$. As described in the previous 
section, a large Sommerfeld enhancement calls for large $f_{S\phi}$ coupling. We can therefore constraint 
the parameter space yielding such a large boost factor with the direct detection bounds. As shown in Fig.~(\ref{fig-5}), a large portion of the green (gray) region is excluded, but nonetheless a large portion is found compatible 
with direct detection constraints. We also show the prediction for the upcoming XENON100 first run (blue line): while 
it can probe a bigger portion of the hidden sector parameter space compatible with indirect detection, a 
large part can give DM to nucleon cross-section below the expected sensitivity.
\begin{figure}[t!]	   
\includegraphics[width=0.4\textwidth]{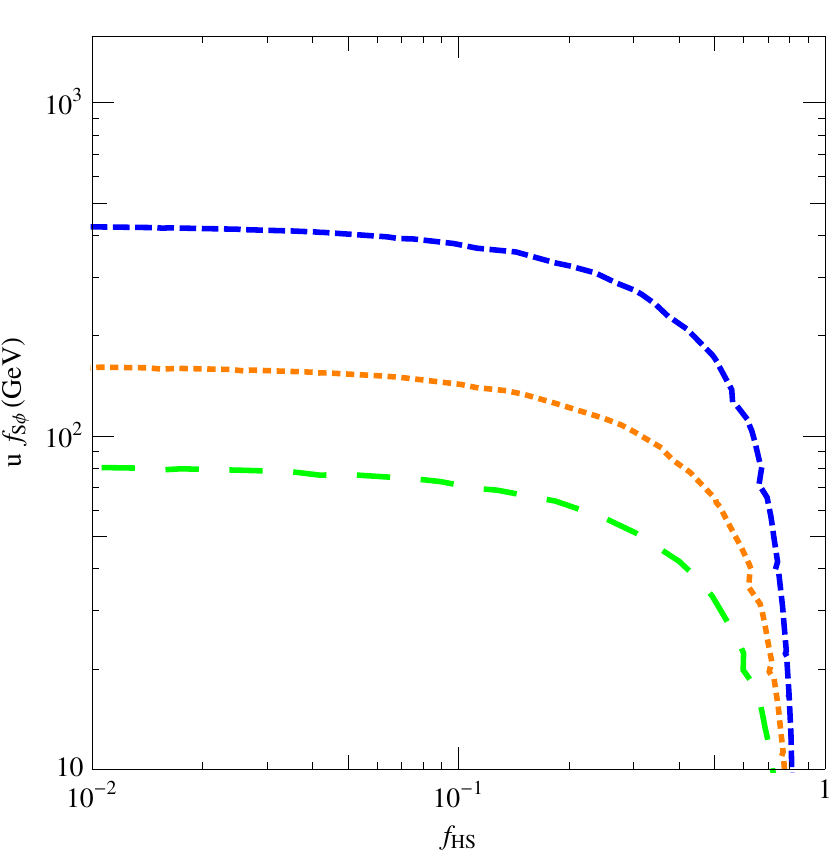} \hspace{2cm} \includegraphics[width=0.4\textwidth]{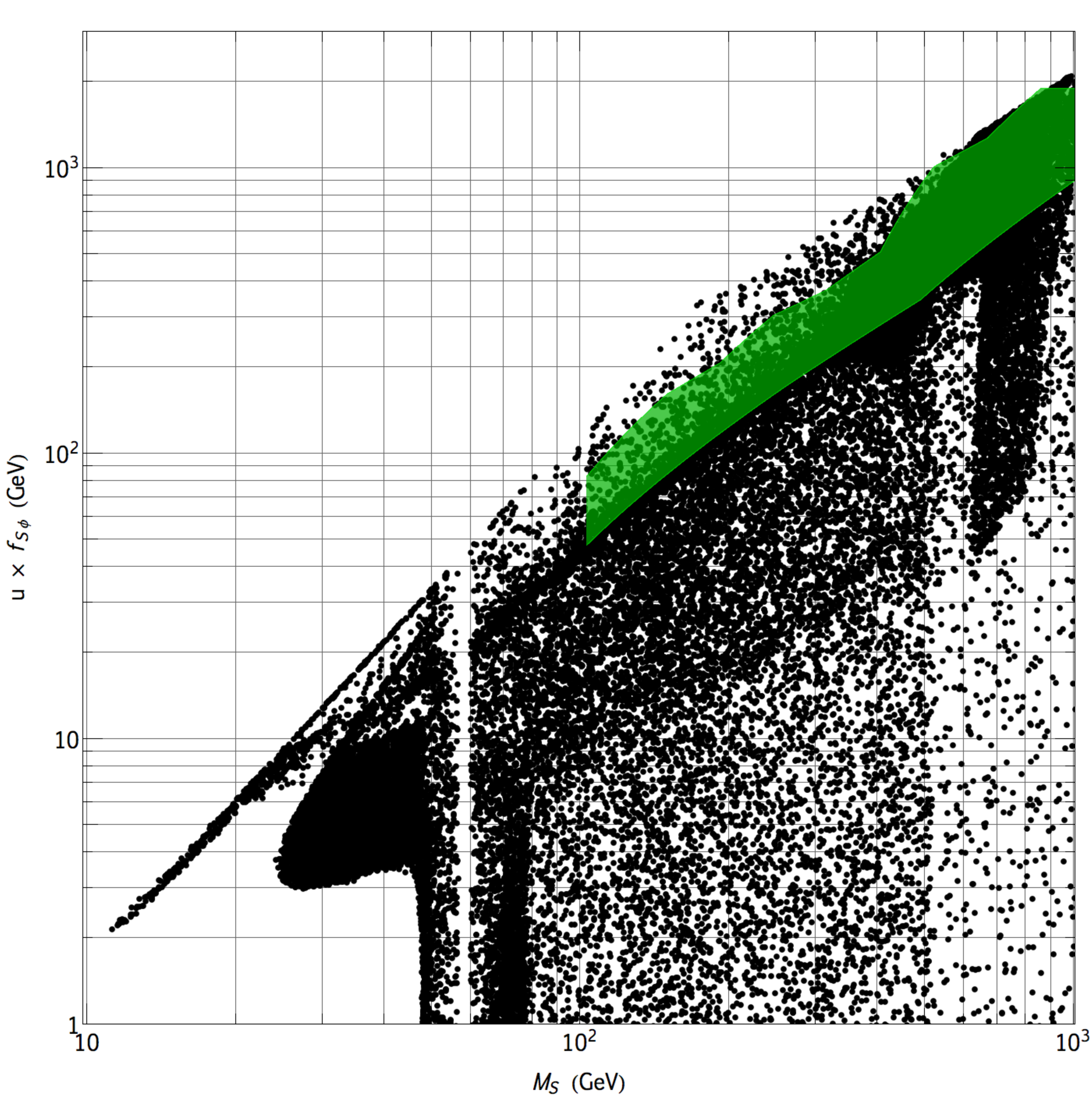}
  \caption{Left panel: maximal $\sigma_{\rm SI}$ value in the plane $f_{HS}$ and $u \times f_{S\phi}$ for $M_S=400$ GeV, $\theta_m = 1.9\times 10^{-6}$ (blue dashed curve), $\theta_m=5\times 10^{-6}$ (dotted orange line) and $\theta_m= 10^{-5}$ (green long dashed line). The region plane on the right-hand side of each curve is excluded by CDMS-II at $90\%$ C.L.. Right panel: scatter plot in the plane of $u\times f_{S\phi}$ vs $M_S$. Black points give the required relic density, while the points in the green (gray) region give rise to a boost factor compatible with PAMELA anomaly.}
\label{fig-4}
\end{figure}

On the left of Fig.~(\ref{fig-4}), we show an illustrative example of the balance between the two contributions in 
Eq.~(\ref{eq:sigmadd}), for a fixed DM mass $M_S=400\GeV$.
The maximum value of the elastic cross section compatible with 
CDMS-II is plotted, as a function of $f_{HS}$ and $u \times f_{S\phi}$. In this plot, three different values of the mixing angle are depicted, $\theta_{m}=1.9\times 10^{-6}$ (blue dashed line),  $5\times 10^{-6}$ (orange dotted line), and $10^{-5}$ (green long dashed line).
First of all, for $f_{HS}\sim 1$, the SM Higgs exchange dominates and $\sigma_{\rm SI}$ is rather insensitive to $\theta_{m}$. As $f_{HS}$ decreases towards smaller values, the $h_{2}$ contribution becomes the leading one. For $f_{HS}\ll 1$, from Eq.~(\ref{eq:sigmadd}) we see that $\sigma_{\rm SI}$ only depends on the product $\theta_{m}\times u\times f_{S\phi}$. For $f_{HS}\sim 10^{-2}$, we then have that different values of the mixing angle imply different maximum values of $u\times f_{S\phi}$.
Notice that, from Eqs.~(\ref{SomPot}) and (\ref{IntSom}), the Sommerfeld enhancement is sensitive to the same combination $u\times f_{S\phi}$.
The right panel of Fig.~(\ref{fig-4}) shows the points giving the right relic abundance in the plane $u\times f_{S\phi}$ vs $M_S$, where again a green (gray) region highlights the values of interest for the indirect detection. For $f_{HS}= 10^{-2}$, a small mixing angle $\theta_{m}=1.9\times 10^{-6}$ allows for $u\times f_{S\phi}\simeq 410\GeV$, which is compatible with indirect detection constraints. Indeed, we see that for $M_S=400\GeV$, $u\times f_{S\phi}$ should be in the $300-500 \GeV$ range. Increasing $\theta_{m}$ calls for smaller values of $u\times f_{S\phi}$, and consequently lower $\langle S_{e} \rangle$ can be obtained. For $\theta_{m}=5\times 10^{-6}$, we see that $u\times f_{S\phi}\lesssim 160 \GeV$, a value for which the boost factor is too small to account for the whole cosmic ray excesses. The situation is even worst for $\theta_{m}=10^{-5}$ for which $u\times f_{S\phi}\lesssim 80 \GeV$.
\section{Electron, Positron and Antiproton Fluxes from $S^\dagger S$ Annihilation}\label{sec:ind}
As described in Sec.~\ref{sec:omegah2}, the annihilation of $S^\dagger S \to h_{2} h_{2}$ and 
$S^\dagger S \to h_{1} h_{1}$ will generate both positrons and antiprotons in the 
present Universe and could have been detected by various experiments such as PAMELA, 
HESS and FermiLAT. In the case $S$ annihilates into $h_{2}$, the final products are dominantly muons and 
antimuons resulting from the decay of $h_{2}$, while in the latter case the final 
products are mostly hadrons. From Eq.(\ref{CrossSections}), the ratio of $S$ annihilation cross sections into $h_{1}$ to $h_{2}$  is 
\begin{equation}
\frac{\langle \sigma \vert v_{\rm rel}\vert \rangle(S^\dagger S \to h_1h_1)}{\langle \sigma \vert v_{\rm rel}\vert \rangle(S^\dagger S \to h_2h_2)} \approx \left( \frac{f_{HS}}{f_{S\phi} } \right)^2\,.
\end{equation}  
As we see from Fig.(\ref{fig-2}), we typically have $f_{S\phi} \gsim 10 f_{HS}$ 
when we require that the annihilation cross-section is enhanced by a non-negligible boost 
factor. This is sufficient to suppress the antiproton flux over the positron one. Therefore, in what follows we will focus on the production and propagation of 
positrons in the Galactic medium.
\subsection{Production and propagation of positrons}
From $S$ annihilations, $h_{2}$ is produced which then decays to muon and antimuon. They 
ultimately decay to electrons, positrons and neutrinos. As a result, equal numbers of electrons and 
positrons are produced from the annihilation of $S$ particles. However, the background 
flux of electrons in the Galactic medium is significantly larger than the positron one. Therefore, 
it is easier to find signature of DM, if any, in the Galactic positron flux. 

Once the positrons are produced in the Galactic halo where the DM concentration is large, 
they travel under the influence of the Galactic magnetic field which is assumed to be of the order of 
a few microgauss. The motion of positrons can then be thought of as a random walk. In the vicinity of the Solar System, the positron flux can be obtained by solving the diffusion 
equation~\cite{delahayeetal:2007}
\bea
\frac{\partial }{\partial t} f_{e^+}(E,\vec{r},t) &=&Q(E,\vec{r}) + K_{e^+}(E) \nabla^2 f_{e^+}(E,\vec{r},t) +
\frac{\partial}{\partial t}[b(E) f_{e^+}(E,\vec{r},t)] \,,
\label{diffusion}
\eea
where $f_{e^+}(E,\vec{r})$ is the number density of positrons per unit energy,
$E$ is the energy of positron, $K_{e^+}(E)$ is the diffusion constant, $b(E)$ is the 
energy-loss rate and $Q(E,\vec{r})$ is the positron source term. The latter, due to $S$ annihilations, is given by:
\be
Q(E,\vec{r})=n_S^2(\vec{r}) \langle \sigma |v_{\rm rel}|\rangle \frac{d N_{e^+}}{d E}\,.
\label{source}
\ee
In the above equation the fragmentation function $d N_{e^+}/d E$ represents the
number of positrons with energy $E$ which are produced from the annihilation of
$S$ particles. We assume that the positrons are in steady state, i.e. 
$\partial f_{e^{+}}/\partial t=0$. Then from Eq. (\ref{diffusion}), the positron flux in 
the vicinity of the solar system can be obtained in a semi-analytical 
form~\cite{delahayeetal:2007,hisanoetal:PRD2006,cireli&strumia:NPB2008}
\bea
\Phi_{e^+} (E,\vec{r}_{\odot}) & = & \frac{v_{e^+}}{4\pi b(E)}(n_S)_{\odot}^2 
\langle \sigma |v_{\rm rel}| \rangle \int_E^{M_S} dE' \frac{dN_{e^+}}{dE'}I (\lambda_D(E,E'))\,,
\label{positron_flux}
\eea
where $\lambda_D(E,E')$ is the diffusion length from energy $E'$ to energy $E$ 
and $I(\lambda_D(E,E'))$ is the halo function which is independent of particle physics. An 
analogous solution for the electron flux can also be obtained.
\subsection{Background fluxes of electron and positron}
Positrons in our galaxy are not only produced by $S$ particle annihilations but 
also by the scattering of cosmic-ray protons with the interstellar 
medium~\cite{moskalenko&strong:astro1998}. The positrons produced from the 
later sources thus act as background for the positrons produced from DM annihilations.
The background positron fraction can be defined as
\be
\left( \frac{\Phi_{e^+}}{\Phi_{e^+}+\Phi_{e^-}} \right)_{\rm bkg} = \frac{\Phi_{\rm sec,\; e^+}^{\rm bkg}}
{\Phi_{\rm prim,\; e^-}^{\rm bkg} + \Phi_{\rm sec,\; e^-}^{\rm bkg}
+ \Phi_{\rm sec,\; e^+}^{\rm bkg}}\,,
\ee
where the primary and secondary electron fluxes, as well as the secondary positron flux,  
can be parameterized as~\cite{baltz&edsjo:prd1998}:
\bea
\Phi_{\rm prim,\; e^-}^{\rm bkg}& =& \frac{0.16 \epsilon^{-1.1}}{1+11 \epsilon^{0.9}+3.2\epsilon^{2.15}}
{\rm GeV^{-1} cm^{-2} s^{-1} sr^{-1}}\,,\nonumber \\
\Phi_{\rm sec,\; e^-}^{\rm bkg} &= &\frac{0.70\epsilon^{0.7}}{1+11\epsilon^{1.5}+600 \epsilon^{2.9}
+580\epsilon^{4.2}} {\rm GeV^{-1} cm^{-2} s^{-1} sr^{-1}}\,,\nonumberÊ\\
\Phi_{\rm sec,\; e^+}^{\rm bkg} &=& \frac{4.5 \epsilon^{0.7}}{1+650 \epsilon^{2.3}+1500 \epsilon^{4.2}}
{\rm GeV^{-1} cm^{-2} s^{-1} sr^{-1} }\,,
\eea
where $\epsilon$=E/(1 GeV) is a dimensionless parameter.
\subsection{Results and Discussions}
\begin{figure}[t!]
\begin{center}
\includegraphics[scale=0.45]{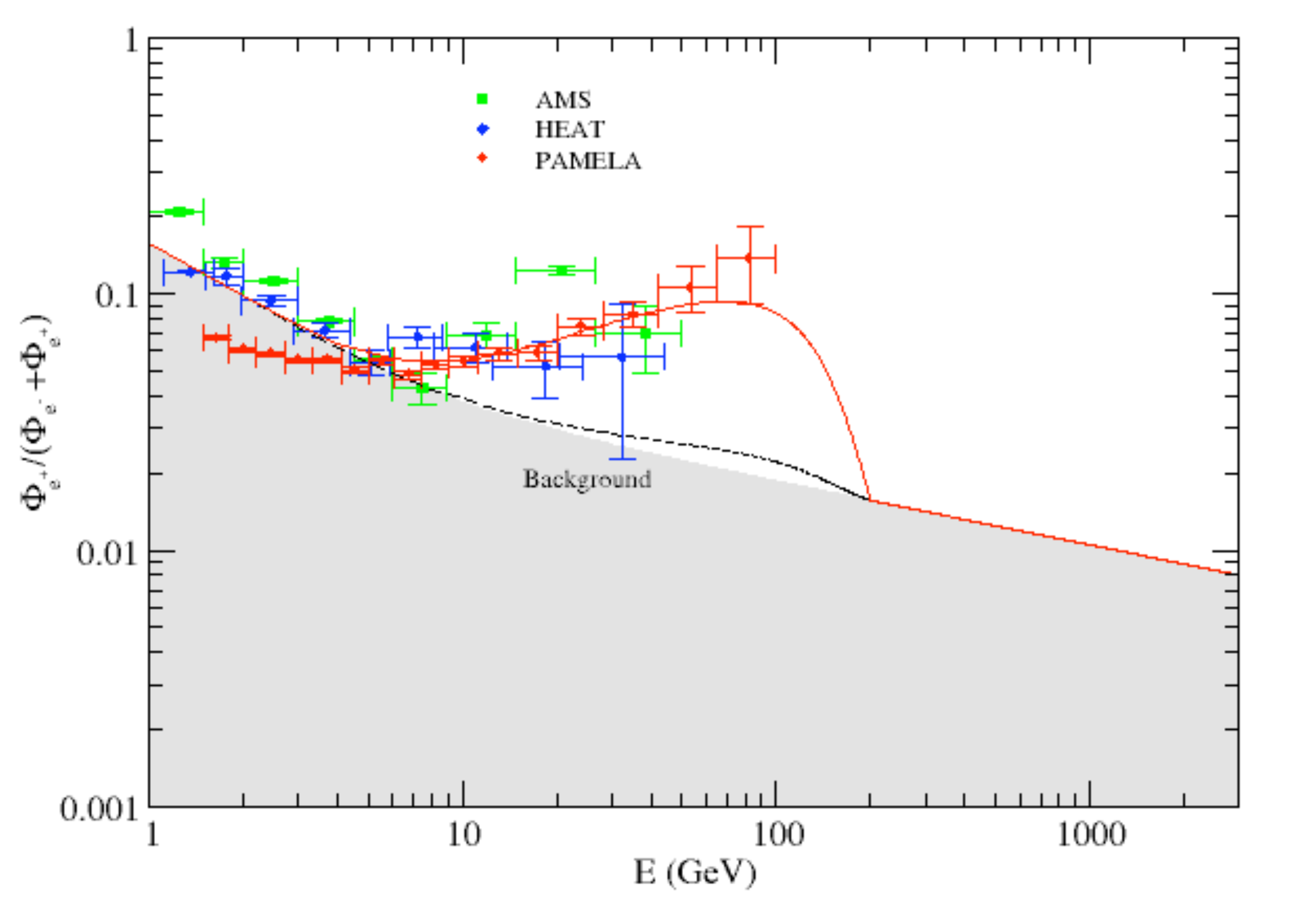}
\caption{Positron fraction from $S^\dagger S \to 2\mu^+\mu^-$ at $M_S=400$ GeV. The red solid line (black dashed line) stands for the illustrative point 1 (2) quoted in Table~\ref{bench}.}
\label{pamela_fitting}
\end{center}
\end{figure}
The net positron flux in the galactic medium is given by
\be
(\Phi_{e^+})_{\rm Gal}=(\Phi_{e^+})_{\rm bkg} + \Phi_{e^+}(E,\vec{r}_{\odot})\,.
\ee
The second term in the above equation is given by Eq. (\ref{positron_flux}), which depends 
on various factors: $b(E)$, $\lambda_D(E,E')$, $I (\lambda_D(E,E'))$, $v_{e^+}$, $(n_S)_{\odot}$ 
and the injection spectrum $dN_{e^+}/dE'$. The energy loss (due to inverse Compton scattering 
and synchrotron radiation with Galactic magnetic field) term $b(E)$ is determined by the photon 
density, and the strength of magnetic fields. Its value is taken to be $b(E)=10^{-16} \epsilon^2 
{\rm GeV s}^{-1}$~\cite{baltz&edsjo:prd1998}. The number density of $S$ DM in the Solar 
System is given by $(n_S)_{\odot}=\rho_\odot/M_S$, where $\rho_\odot\approx 0.3 {\rm GeV/cm^3}$. 
In the energy range we are interested in, the value of $v_{e^+}$ is taken approximately to be $c$, 
the velocity of light. The values of diffusion length $\lambda_D(E,E')$ and the corresponding halo 
function $I (\lambda_D(E,E'))$ are based on astrophysical assumptions~\cite{delahayeetal:2007,
cireli&strumia:NPB2008}. By considering different heights of the Galactic plane and different DM 
halo profiles the results may vary slightly. In the following, the Galactic plane height is taken to be less than
$4$ kpc, which is referred to as the "MED" model~\cite{delahayeetal:2007,cireli&strumia:NPB2008}, 
and we have used the Navarro-Frenk-White (NFW) DM halo profile~\cite{NFW}
\be
\rho(r)=\rho_{\odot}\left( \frac{r_\odot}{r} \right)\left(\frac{1+ \left( \frac{r_\odot}{r_s} \right)
}{1+\left(\frac{r}{r_s} \right)} \right)^2
\label{NFW-profile}
\ee
to determine the halo function $I (\lambda_D(E,E'))$, where $r_s\approx 20 {\rm kpc}$ and 
$r_{\odot} \approx 8.5 {\rm kpc}$. 

We use the program DARKSUSY~\cite{darksusy} to compute electron and positron fluxes from $S$ annihilations $S^{\dagger}S\to h_{2}h_{2},h_{2}\to \mu^{+}\mu^{-}$. We then determine, for $100\GeV \lesssim M_S\lesssim 1\TeV$, what is the maximum annihilation cross section allowed for the fluxes not to exceed PAMELA observations. In this range, we found the approximate empirical upper bound:
\be
\langle S_e \rangle \lesssim 1800\times\left(\frac{M_S}{1\TeV}\right)^{1.95}\,.
\label{BSbf}
\ee
The constraint Eq.~(\ref{BSbf}) only tells us that bigger boost factor are excluded. If one wants to fully account for the anomalous cosmic ray fluxes through DM annihilations, the boost factor gets lower-bounded, again in the $100 \GeV \lesssim M_S \lesssim 1 \TeV$ range:
\bea
\langle S_e \rangle \gtrsim 1000\times\left( \frac{M_S}{1\TeV}\right)^{1.85}\,.
\label{BIbf}
\eea
If both Eqs.~(\ref{BSbf}) and (\ref{BIbf}) are fulfilled, then cosmic ray fluxes are within 1 standard deviation of PAMELA data. 
Of course, if all PAMELA data have to be explained, $M_S\gtrsim 400\GeV$ is required, given the annihilation channel $S^{\dagger}S\to h_{2}h_{2},h_{2} \to \mu^{+}\mu^{-}$.

Actually several constraints exist on large annihilation cross sections, relying on different physics, but all sensitive to DM annihilation products. When compared to the fiducial value of the annihilation cross section, these constraints apply in turn to the Sommerfeld enhancement $\langle S_e \rangle$.
At high redshift, the energy deposition of the charged leptons may induce perturbations of the cosmic microwave background photon spectra~\cite{CMB1}, reionization and heating of the intergalactic medium~\cite{reion}, providing strong constraints. At the recombination time during which DM relative velocity is $\beta\sim 10^{-8}$, a bound on $S_e$ is inferred~\cite{CMB1}:
\be
S_{e}\lesssim 480\times\left(\frac{M_{S}}{1\TeV}\right)\,.
\label{BScmb}
\ee 
Stringent constraints arise from inverse Compton gamma rays in the Galaxy. The muons produced in DM annihilations subsequently decay into electrons. This population of electron yields irreducible high-energy gamma rays through inverse Compton on the Galactic radiation field. We consider the Fermi data released in \cite{Abdo:2010nz}, at galactic latitude $|b|>10^{\circ}$ and the analysis of Ref.~\cite{Hutsi:2010ai}. For a NFW profile and a final state into $4\mu$, the allowed boost factor is $\sim 300$ for a dark matter mass of 400 GeV. Notice that Ref.~\cite{Papucci:2009gd} considers a Galactic latitude closer to the Galactic center and is therefore more sensitive to the DM density profile. In this case and for a NFW profile the maximum allowed boost factor at $M_{S}=400 \GeV$ is $\sim 100$. The model is thus in great tension with the Pamela anomaly. However with an isothermal profile instead, a boost factor up to $\sim 1000$ is allowed, attenuating the constraints on the model parameter space. As for the extragalactic gamma ray constraints, discussed in Refs.~\cite{Abdo:2010dk,Hutsi:2010ai}, they strongly depend on the assumptions on the history of structure formation. It turns out that the parameter space we are considering is allowed for a conservative choice of the halo concentration parameter, see Fig. 4 of \cite{Hutsi:2010ai} and Fig. 6 of Ref.\cite{Abdo:2010dk} (which considers a two-muon final state case). For the cases we consider, constraints from recombination and from diffuse gamma rays are of the same order. In Figs.(\ref{fig-2})-(\ref{fig-4}), we depicted the boost factor satisfying the constraints Eqs.(\ref{BSbf}-\ref{BScmb}) within a green (gray) band.

\begin{table}[t]
  \begin{center}
    \begin{tabular}{c|ccccc}
      \hline\hline
      Point & $f_{S\phi}$ & $f_{HS}$ & $u$ & $\theta_m$ & $\langle S_e \rangle$\\
      \hline\hline
      1 & 0.82  & $10^{-2}$ & 500  & $1.9\times 10^{-6}$ & 195 \\
       2 & 0.1 & $10^{-2}$ & 800  & $5\times 10^{-6}$ & 8  \\
      \hline\hline
    \end{tabular}
 \caption{Typical points used for the analysis of the PAMELA positron fraction. The DM mass is fixed at 400 GeV.}
       \label{bench}
  \end{center}
\end{table}

In Fig.~(\ref{pamela_fitting}), we show the comparison between the positron fraction obtained from 
$S$ annihilations with the positron fraction observed by PAMELA, AMS and HEAT, for the two typical points defined in Table~\ref{bench} at $M_S=400$ GeV. The quoted values of the couplings are inferred from Fig.~(\ref{fig-4}), and provide a relic density in the WMAP7 range as well as saturate current direct detection bound. The first set of parameters is in a good agreement with the anomalous positron fraction observed by PAMELA. However, the corresponding boost factor $\langle S_e \rangle \sim 200$ is at the border line of gamma ray constraints coming from FermiLAT~\cite{Hutsi:2010ai}-\cite{Abdo:2010dk} and reionization~\cite{reion}. The second set cannot entirely account for the excess measured by PAMELA since the largest boost factor allowed by direct detection is about 8. We conclude that in order to fully explain the positron fraction together with an observable cross section on nucleon, bigger values of $u\times f_{S\phi}$, and conversely lower values of the mixing angle are mandatory. 
\begin{figure}[ht!]
\begin{center}
\includegraphics[scale=0.47]{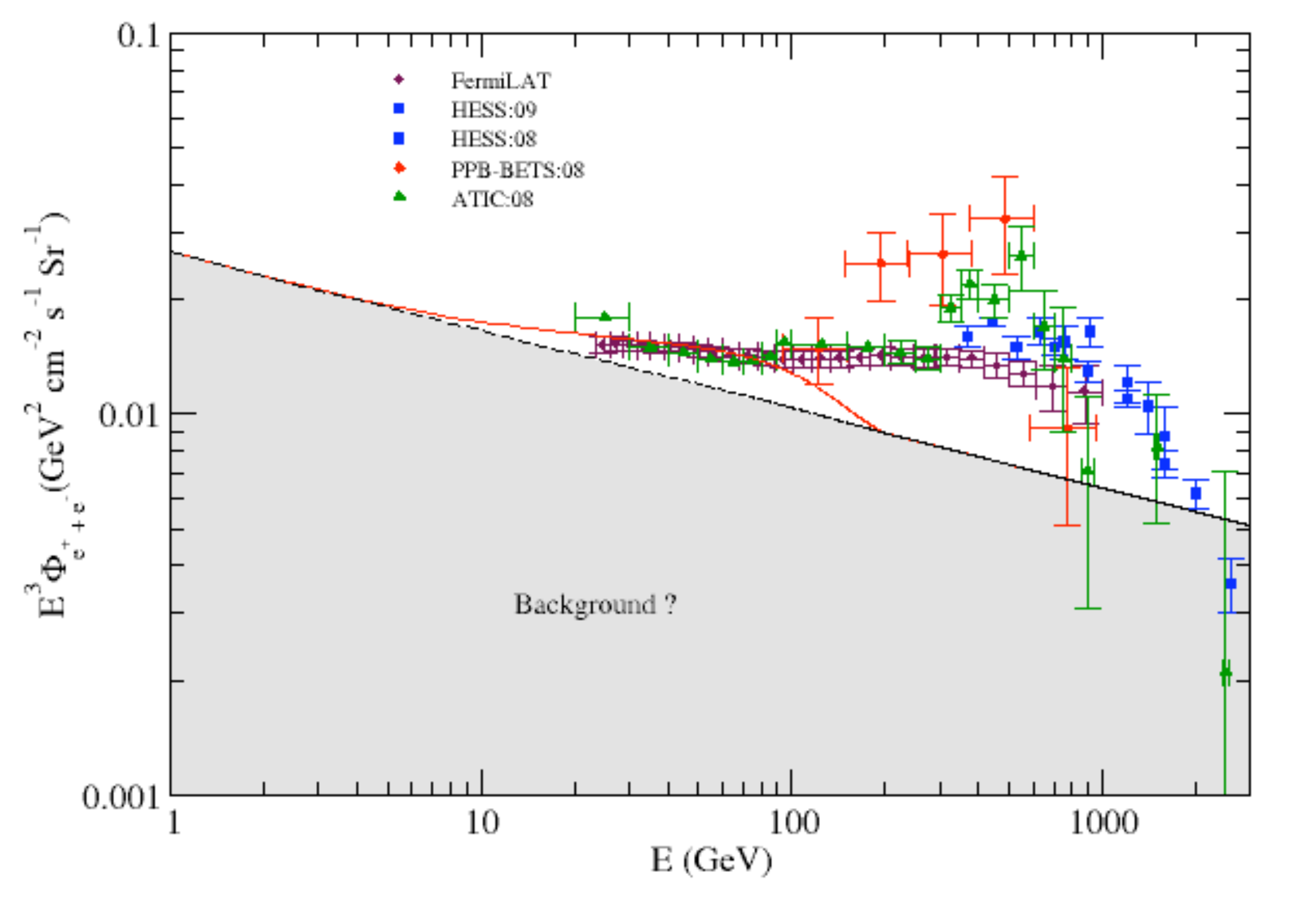}
\caption{Electron plus positron flux from $S^\dagger S \to 2\mu^+\mu^-$ at $M_S=400$ GeV. The red solid line (black dashed line) stands for the illustrative point 1 (2) quoted in Table~\ref{bench}.}
\label{fermi_fitting}
\end{center}
\end{figure}

Concerning the electron plus positron flux observed by FermiLAT, $M_S\sim 1\TeV$ is necessary if the flux stems only from DM annihilations. However, for such high masses, the required boost factor is  $\langle S_e \rangle \gtrsim 10^{3}$. Such a high value is in great tension with reionization constraints~\cite{reion}. 
In Fig.~(\ref{fermi_fitting}), we compare FermiLAT, HESS, PPB-BETS and ATIC data on electron plus positron fluxes with the predictions of this model for the representative points mentioned in Table~\ref{bench}, for which $M_S=400\GeV$. The first point, depicted by a red solid line, yields $\langle S_e \rangle =195$ and can fit PAMELA data as well as the low energy flux observed by FermiLAT. Of course, with $M_S=400 \GeV$, not all the energy range can be explained. The second point, depicted by a dashed black line and for which $\langle S_e \rangle =8$, cannot account neither for the PAMELA results nor for the FermiLAT ones. In both cases, FermiLAT electron plus positron flux can only explained by adding an extra source of astrophysical origin. 
\section{Conclusions}\label{sec:conclusion}
In this paper we studied a hidden Abelian extension of the standard model. The DM is a 
complex scalar, singlet under the SM gauge group but charged under the hidden sector. We also 
introduce a light scalar $\Phi$, whose non-zero vev $u$ breaks $U(1)_{H}$ to a $Z_2$ symmetry 
under which all fields but $S$ are even. As a result $S$ is a stable DM candidate, with mass ranging from the GeV to the TeV scale. The three fields $H$, $\Phi$ and $S$ couple together via three Higgs portal couplings. The physical scalars $h$ and $\phi$ mix together with the mixing angle $\theta_{m}\propto f_{H\phi}$.

The relic density of $S$ mainly results from the annihilation channel $S^{\dagger}S\to h_{2} h_{2}$ through $f_{S \phi}$. 
All three portal couplings enter in the spin-independent DM-nucleon cross section. While the usual Higgs-mediated channel depends on $f_{HS}$, the mixing between $h$ and $\phi$ provides an additional channel, mediated by the light scalar $h_{2}$, which is $\propto u\times f_{S\phi}\times \theta_{m}$.
Given the mass scales we consider, the main contribution to the direct detection signal actually comes from this mixing term. For the 
parameter space we scanned over, solutions are found saturating or exceeding current experimental bounds from XENON10 and CDMS-II, or are in the reach of sensitivity of 
XENON100. 
The model also provides indirect signatures of $S$ through cosmic ray flux measurements. The main annihilation 
channel $S^{\dagger}S\to h_{2} h_{2}$, followed by fast $h_{2}\to \mu^{+}\mu^{-}$ decays, ends up in 
high-energy electron and positron fluxes. The suppression of $S$ annihilations into standard Higgs 
compared to our dominant channel, together with the light mass of $h_{2}$, entail that no antiproton flux is expected at a significant level.
A Sommerfeld enhancement of the current DM annihilation cross section occurs through the light $h_{2}$ exchange, which explains cosmic ray excess observations. This enhancement depends on $u\times f_{S\phi}$.

Interestingly, in this model, direct and indirect DM searches constrain same part of the parameter space.
More precisely, in order to fully account for the anomalous positron fraction observed by PAMELA, large values of $u\times f_{S\phi}$ are required. Such large values give rise to large direct detection signals, saturating the current experimental exclusion limits. From this, an upper bound on the mixing angle is inferred. As an example, for $M_S=400\GeV$ and $u\times f_{S\phi}\sim 410 \GeV$, $\theta_{m}$ should be less than $1.9\times 10^{-6}$ in order to satisfy simultaneously the current direct and indirect detection limits.
We stress that in all models where the Sommerfeld enhancement occurs thanks to a light scalar that mixes with the Higgs particle, direct and indirect detection of dark matter are tightly connected.

\acknowledgments
N.S. would like to thank K. Kohri, J. McDonald and C. Balazs for useful discussions.
This work is supported by  the IISN and the Belgian Science Policy (IAP VI-11).

\end{document}